\documentclass[twocolumn,showpacs,prb]{revtex4}
\usepackage{hyperref}
\usepackage{amsmath,amssymb}
\usepackage{epsfig}
\usepackage{graphicx}
\usepackage{graphics}
\usepackage{dcolumn}

\begin{document}

\title{Frequency-dependent ratchet effect in superconducting films with a tilted washboard pinning potential}

\author{Valerij~A.~Shklovskij$^{1,2}$ and Oleksandr~V.~Dobrovolskiy$^3$}

\address
        {$^1$Institute of Theoretical Physics, NSC-KIPT, 61108 Kharkiv, Ukraine\\
        $^2$Physical Department, Kharkiv National University, 61077 Kharkiv, Ukraine\\
        $^3$Physikalisches Institut, Goethe-University, 60438 Frankfurt am Main, Germany}

\date{\today}

\begin{abstract}
The influence of an ac current of arbitrary amplitude and frequency on the mixed-state dc-voltage-ac-drive ratchet
response of a superconducting film with a dc current-tilted uniaxial cosine pinning potential at finite temperature
is theoretically investigated. The results are obtained in the single-vortex approximation, i.e., for non-interacting vortices,
within the frame of an exact solution of the appropriate Langevin equation  in terms of a matrix continued fraction. Formulas
for the dc voltage ratchet response and absorbed power in ac response are discussed as functions of ac current amplitude and
frequency as well as dc current induced tilt in a wide range of corresponding dimensionless parameters. Special
attention is paid to the physical interpretation of the obtained results in adiabatic and high-frequency ratchet
responses taking into account both running and localized states of the (ac+dc)-driven vortex motion in a washboard
pinning potential. Our theoretical results are discussed in comparison with recent experimental work on the high-frequency
ratchet response in nanostructured superconducting films [B.~B. Jin \emph{et al.}, Phys. Rev. B \textbf{81}, 174505 (2010)].
\end{abstract}

\pacs{74.25.F-, 74.25.Wx, 74.25.Qt, 74.40.De}

\maketitle

\section{INTRODUCTION}
Within the last decade vortex ratchets, which exploit asymmetric vortex dynamics, have been attracting considerable
attention~\cite{Reimann}-\cite{jin10}. In essence, the vortex ratchet is a system where the vortex can acquire a net motion in
an asymmetric periodic pinning potential (PPP) in the presence of deterministic or stochastic forces with time averages of
zero~(for comprehensive reviews, see~\cite{Reimann}-\cite{han09}). The asymmetry in the PPP refers to the current direction
reversal. There are essentially two different ways to realize such a pinning potential asymmetry. First, the spatial inversion
symmetry of the PPP itself can be
broken intrinsically and involves some kind of periodic and asymmetric pinning potential, also known as a \emph{rocking ratchet}.
A second option is that an initially symmetric PPP, if externally biased, i.e., subjected to an additive constant driving force,
results in an effective asymmetric pinning potential. This is called a tilted-potential ratchet or a \emph{tilting ratchet}.
Irrespective of the way to bring the asymmetry into a system, from a practical viewpoint, the common feature of superconducting
ratchets is their rectifying property: the application of an ac current to a superconductor with an asymmetric PPP landscape can
produce vortex motion whose direction is determined only by the asymmetry of the pinning potential.

A considerable amount of theoretical work about the general properties of different types of ratchet systems exists~\cite{Reimann,han09,shk09}.
Such ratchet systems range from the use of Josephson junctions in superconducting
quantum interference devices (SQUIDs) and arrays~\cite{zap96}, to the use of one- or two-dimensional
potential-energy ratchets allowing one to construct fluxon pumps and lenses~\cite{wam99}, and drive fluxons out of superconducting samples~\cite{Lee}.
Whereas the majority of ratchet proposals rely on single particles interacting with an external potential to produce the dc response,
collective interactions between particles needed to produce dc transport have also been considered previously~\cite{ols01}. At the same
time in experiments, initially asymmetric PPPs have been used rather than dc-biased ones so far. For instance, the vortex lattice ratchet
effect has been investigated in Nb films sputtered on arrays of nanometric Ni triangles, which produce the asymmetric PPP~\cite{vil05}.
Similar effects were also discussed for YBCO films with antidots~\cite{wor04}. The voltage rectification in superconducting Al films
patterned with either asymmetric or symmetric antidots and subjected to an ac driving current, dc-biased in the symmetric case, has
been experimentally observed in Ref.~\cite{sil06}. Among other experimental works on the vortex ratchet effect in nano-patterned
superconductors, two substantial recent papers~\cite{per09}~\cite{jin10} should be mentioned. D.~Perez de Lara \emph{et al.}~\cite{per09}
have investigated ratchet effects in thin Nb films grown on top of arrays of Ni nanotriangles subjected to an ac current with a
frequency up to 10 kHz, so that effects observed in that work~\cite{per09} were adiabatic, i.e., independent on ac frequency.
Only recently, B.~B. Jin \emph{et al.}~\cite{jin10} have experimentally investigated a very important issue in the vortex
ratchet study, namely the frequency dependence of the dc voltage at large amplitudes of the ac driving force in a frequency
range between 0.5~MHz and 2~GHz. As it was pointed out in Ref.~\cite{jin10}, ac frequencies were always lower than 1 MHz in
vortex ratchet measurements up to that work~\cite{jin10} (see Refs.~[14-20] therein).

So far, a full temperature-dependent theoretical description of the superconducting devices proposed in Refs.~[9-13] is not
available due to the complexity of the two-dimensional PPP used in these references. In particular, the theoretical explanation
of the experimental results of the vortex flow along the vortex channeling directions in the above-mentioned structures is a
difficult problem. Below we propose to experimentally study ratchet properties on nanostructured thin-film superconductors~\cite{dob10-1,dob10-2}
with an uniaxial, i.e., \emph{washboard} pinning potential (WPP). Uniaxiality of the proposed potential does not mean that the physics
of vortex motion in a symmetric WPP, tilted by an external dc bias, becomes one-dimensional as the angular component of the
moving force could simply be regarded as changing the strength of the WPP. Besides the two limiting cases of transversal and
longitudinal geometry, when the vortices move across or along the WPP channels, respectively, one has to consider for all
intermediate angles not only longitudinal, but also transverse ratchet effect~\cite{shk09,per09}, where the last appears
due to the guided vortex motion along the WPP channels.

In the present work, however, we will use the transverse geometry only to provide the reader with most intuitive figure data
and to simplify the subsequent analysis. Whereas the model allows one to obtain exact expressions for the magneto-resistivities
at any intermediate current angles with regard to the WPP channels, general formulas for the responses have been provided
recently~\cite{shk08}. In that work~\cite{shk08}, a full and exact theoretical description of the nonlinear in current and
temperature vortex dynamics in the ratchet devices have been performed in the single-vortex approximation, i.e., for non-interacting
vortices, within the framework of the Langevin equation. It should be noted that theoretical predictions of Ref.~\cite{shk08}
lack experimental scrutiny so far, especially at microwave and GHz frequencies. For this reason, throughout this paper
where applicable, we will be referring to the results of B.~B. Jin \emph{et al.}~\cite{jin10} as a most closely related
experimental work to our new tilted-ratchet results, though that work~\cite{jin10} is dealing with a rocking ratchet.

In more detail, the mixed-state resistive response of superconducting films has been theoretically investigated in Ref.~\cite{shk08}
in the high frequency and strong amplitude regime of the ac vortex transport in the presence of a dc bias, which invokes a definite tilt
of the cosine pinning potential, taking nonzero temperature fluctuations also into account. The exact solution of this nonlinear and time-dependent
problem, obtained in Ref.~\cite{shk08} in terms of a matrix continued fraction, included only general formulas for the ac and dc
magnetoresistive responses. These will be elaborated in detail in this paper and applied to the study of both the dc ratchet
electric field response and the absorbed ac power dependences on the ac amplitude and frequency at fixed temperature for arbitrary dc biases
allowing one to adjust the asymmetry of the PPP.

The aim of this paper is to physically analyze the tilted-ratchet problem on the basis of the single-vortex model in order to
determine those "intrinsic" tilted-ratchet effects in the vortex dynamics which arise from the tilt of initially symmetric WPP
as the only reason. In addition, this model allows one to study theoretically the exact ratchet behavior of absorbed power at
strong ac amplitude and arbitrary frequency, i.e., the subject which has not been studied in any known for us previous theoretical
work even for usual vortex ac response. As a result, two groups of new findings have been obtained. Exact formulas for (i) the
dc voltage ratchet response and (ii) absorbed power in ac response will be discussed as functions of ac current amplitude and
frequency as well as dc current induced tilt, in a wide range of corresponding dimensionless parameters. Experimentally,
the obtained results can be verified on superconducting films with a WPP, similar to those used in Refs.~\cite{dob10-1,dob10-2}.
From the viewpoint of basic research, it will be pointed out which new ratchet effects in the vortex dynamics appear even
within the single-vortex approximation. Besides, a further development of the theory towards the consideration of an
asymmetric-potential ratchet will allow one to distinguish tilted-ratchet effects from potential asymmetry-induced
effects when considering an asymmetric WPP in the presence of a dc tilting bias.

The organization of the paper is as follows. In Sec.~II we introduce the model and summarize the expressions for dc and ac ratchet
responses, obtained in terms of a matrix continued fraction. In Sec.~III we graphically analyze these quantities as functions of
their driving parameters, namely dc bias, ac amplitude, and frequency. In Sec.~IV we discuss in detail two limiting cases, the adiabatic
and nonadiabatic regimes, and explain peculiarities in the ratchet responses on the basis of either the static current-voltage
characteristics or solution in terms of the Bessel functions, respectively. In Sec.~V we conclude with a general discussion of
our results outlining the difference between the intrinsic and tilted-ratchet models elucidating their applicability and
drawing parallels between our theoretical results and a recent experiment~\cite{jin10}.

\section{Formulation of the Problem}
\begin{figure}
   \epsfig{file=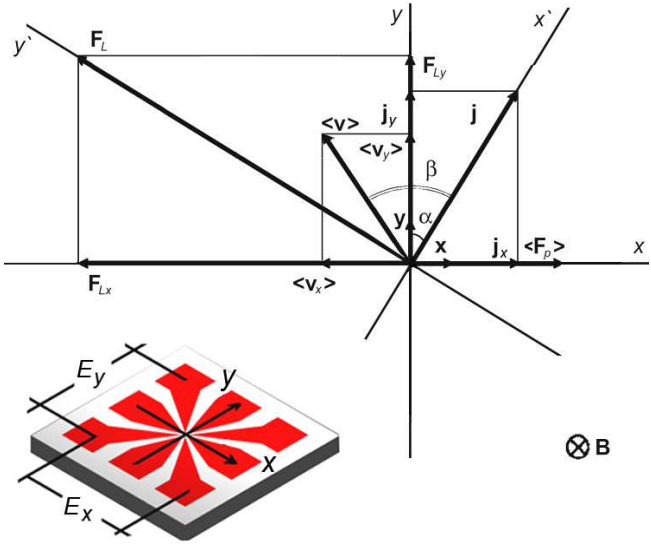,width=8.6cm}
   \caption{The system of coordinates $xy$ with the unit vectors $\mathbf{x}$ and $\mathbf{y}$ is associated with the WPP channels
   which are parallel to the vector $\mathbf{y}$. The coordinate system $x'y'$ is associated with the direction of the transport
   current density vector $\mathbf{j}=\mathbf{j}^{dc}+\mathbf{j}^{ac}\cos\omega t$, $\alpha$ is the angle between $\mathbf{j}$
   and $\mathbf{y}$, $\beta$ is the angle between the average velocity vector $\langle \mathbf{v}\rangle$ and $\mathbf{j}$.
   $\langle \mathbf{F}_p\rangle$ is the average pinning force provided by the WPP, $\mathbf{F}_L$ is the Lorenz force for a vortex,
   and $\textbf{B}$ is the magnetic field vector. Inset: a schematic sample configuration in the general case. Experimentally deducable
   values are the voltages $E_x$ and $E_y$.}
\end{figure}
\begin{figure}
   \epsfig{file=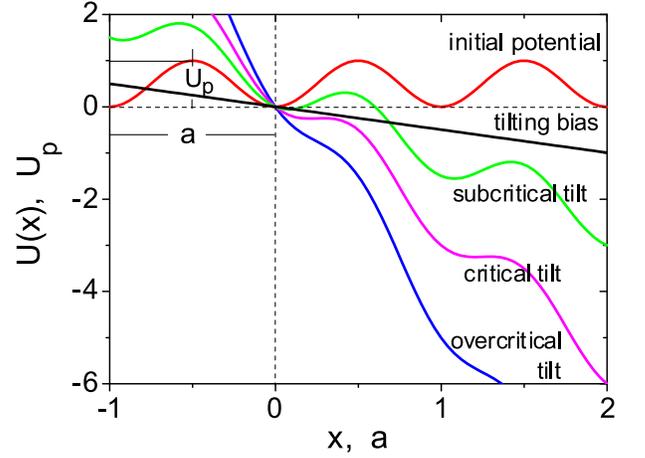,width=8cm}
   \caption{Modification of the effective pinning potential $U(x)\equiv U_p(x) - Fx$ with gradual increase of the Lorentz force
   component in the $x$-direction $F$, where $U_p(x)=(U_p/2)(1-\cos kx)$ is the WPP with its depth $U_p$ and period $a=2\pi/k$.
   As the initial WPP is symmetric, i.e., $U_p(-x)=U_p(x)$, it can establish ratchet properties only in the presence of an external
   dc bias $F$ invoking its tilt. Depending on the bias value, in the absence of an ac current and assuming $T=0$~K for simplicity,
   two qualitatively different modes in the vortex motion appear. (i) If $F < F_p$, thought the initial potential well is tilted,
   it maintains the average vortex position, i.e., the vortex is in the \emph{localized} state. At the critical tilt value, i.e.,
   when $F = F_p$, the right-side potential barrier disappears. (ii) At last, when $F > F_p$, the vortex motion direction coincides
   with the direction of the moving force $F$, i.e., the vortex is in the \emph{running} state with an oscillating instantaneous velocity
   with a frequency $\Omega\approx\sqrt{(\xi^d)^2-1}$ (see Sec.~III for details).}
\end{figure}
Our theoretical treatment of the system, schematically shown in Fig.~1, relies upon the Langevin equation for a vortex moving with
velocity $\mathbf{v}$ in a magnetic field $\mathbf{B}=\mathbf{n}B$ ($B\equiv|\mathbf{B}|$, $\mathbf{n}=n\mathbf{z}$, $\mathbf{z}$~is
the unit vector in the $z$ direction and $n=\pm 1$) which, neglecting the Hall effect, has the form
\begin{equation}
        \label{F1}
        \eta\mathbf{v}=\mathbf{F}_{L}+\mathbf{F}_{p}+\mathbf{F}_{th},
\end{equation}
where $\mathbf{F}_{L}=n(\Phi_{0}/c)\mathbf{j}\times\mathbf{z}$ is the Lorentz force ($\Phi_{0}$ is the magnetic flux quantum, and
$c$ is the speed of light). $\mathbf{j}=\mathbf{j}(t)=\mathbf{j}^{dc}+\mathbf{j}^{ac}\cos\omega t$, where $\mathbf{j}^{dc}$ and
$\mathbf{j}^{ac}$ are the dc and ac current density amplitudes and $\omega$ is the angular frequency. $\mathbf{F}_{p}=-\nabla U_{p}(x)$
is the anisotropic pinning force, $U_p(x)=(U_p/2)(1-\cos kx)$ is the periodic washboard pinning potential with $k=2\pi/a$~\cite{maw99}-\cite{gol92},
where $U_p$ is its depth and $a$ is the period (see Fig.~2). $\mathbf{F}_{th}$ is the thermal fluctuation force and $\eta$ is the vortex viscosity.
We assume that the fluctuational force $\mathbf{F}_{th}(t)$ is represented by a Gaussian white noise, whose stochastic properties are
given by the relations $\langle F_{th,i}(t)\rangle=0$, $\langle F_{th,i}(t)F_{th,j}(t') \rangle=2T\eta\delta_{ij}\delta(t-t')$,
where $T$ is the temperature in energy units, $\langle...\rangle$ means the statistical average, $F_{th,i}(t)$ with $i=x$ or $i=y$
is the $i$ component of $\mathbf{F}_{th}(t)$, and $\delta_{ij}$ is Kronecker's delta.

The Langevin equation~\eqref{F1} with the Hall term has been solved in Ref.~\cite{shk08} in terms of a matrix continued fraction.
Neglecting the Hall effect, which is usually small in conventional type-II superconductors, e.g., in Nb films,
below we summarize only the final expressions needed for the subsequent analysis in Secs.~III and IV, where the
main quantities of physical interest are (i)~the \emph{time-independent} (but frequency-dependent) dc electrical field response and
(ii)~the \emph{stationary} ac response on the frequency $\omega$, independent on the initial conditions. Both these are determined by
the appropriate components of the average electric field induced by the moving vortex system, $\langle \mathbf{E}(t) \rangle$, whose
time-independent $dc$ components, $\langle E^{dc}_x\rangle_0^{\omega}$ and $\langle E^{dc}_y\rangle_0^{\omega}$~\cite{shk08}, are
\begin{equation}
\left\{
    \begin{array}{ll}
    \label{F23}
    \langle E^{dc}_y \rangle_0^{\omega}=n\rho_f j_c(j^{dc}-\langle \sin \textsl{x}\rangle_0^{\omega})
    =\rho_f\nu_0^{\omega}j^{dc}_y\\
    \\
    \langle E^{dc}_x\rangle_0^{\omega}=\rho_fj^{dc}_x.\\
    \end{array}
    \right.
\end{equation}
where $\rho_f\equiv B\Phi_0/\eta c^2$ is the flux-flow resistivity, $j_c\equiv cU_pk/2\Phi_0$, $j^{dc}_y=j^d\cos\alpha$,
$j^{dc}_x=j^d\sin\alpha$, $j^d=|\mathbf{j}^{dc}|$ and $\nu_0^{\omega}$ is the $(\omega, j^{dc}, j^{ac}, T)$-dependent effective mobility
of the vortex under the influence of the dimensionless generalized moving force $j^{dc}\equiv n j^{dc}_y/j_c$ in the $x$ direction being
\begin{equation}
    \label{F24} \nu_0^{\omega}\equiv1-\langle \sin \textsl{x}\rangle_0^{\omega}/j^{dc}.
\end{equation}
The term $\langle\sin \textsl{x}\rangle_0^{\omega}$ represents the time-independent static average pinning force, given by Eq. (24) of Ref.~\cite{shk08}.

The nonlinear power absorption in the ac response per unit volume and averaged over the period of an ac cycle is given in accordance
with Eq.~(85) of Ref.~\cite{shk08} by the following expression
\begin{equation}
    \label{F64}
    \mathcal{\bar{P}}(\omega)=\rho_f/2\cdot(j^{ac})^2[\sin^2\alpha + \cos^2\alpha\textrm{Re}Z_1(\omega)].
\end{equation}
where
\begin{equation}
    \label{F37}
    Z_1(\omega)=1-\langle \sin\textsl{x}\rangle_{t1}/j^{ac}
\end{equation}
is the nonlinear impedance with the term $\langle \sin\textsl{x}\rangle_{t1}$ being the time-dependent dynamic average pinning force,
determined by Eq. (24) in Ref.~\cite{shk08}, $j^{ac}\equiv n j^{ac}_y/j_c$, $j^{ac}_y=j^a\cos\alpha$, $j^{ac}_x=j^a\sin\alpha$, and
$j^a=|\mathbf{j}^{ac}|$.

\section{Main ratchet results}
\begin{figure*}
   \label{e_xi}
   \epsfig{file=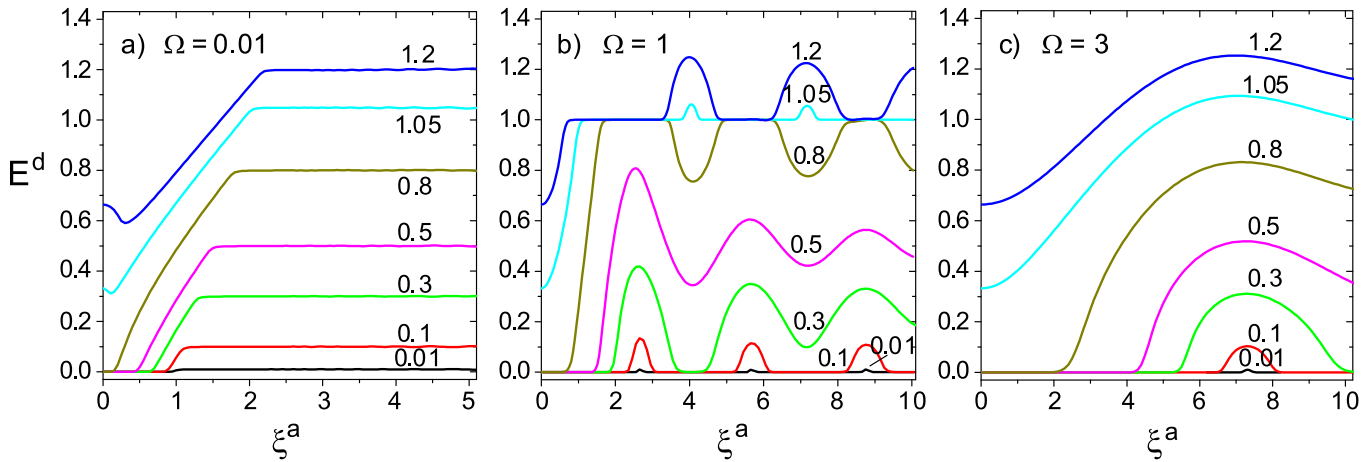,width=18cm}
   \caption{The ratchet voltage $E^d$ versus $\xi^a$ for a set of biases $\xi^d=0.01; 0.1; 0.3; 0.5; 0.8; 1.05; 1.2$, as indicated, in
   the adiabatic~(a), intermediate~(b) and high-frequency~(c) regime.}
\end{figure*}
The main goal of this section is to present results of a detailed theoretical study of the ratchet properties of the (dc+ac)-driven
nonlinear time- and temperature-dependent vortex dynamics within the frames of the exact solution of the problem, presented
recently in our work~\cite{shk08}. Here we study two main physical quantities which can be measured experimentally for
our model, the dc electric field $\mathbf{E}^d$ and the ac power absorption $\mathcal{\bar{P}}$ (in units of $\rho_f$), as
functions of their dimensionless external driving parameters, namely dc bias $\xi^d=j^d/j_c$, amplitude $\xi^a=j^a/j_c$, and
frequency of the ac input $\Omega=\omega\hat{\tau}$ with $\hat{\tau}\equiv2\eta/U_pk^2$ being the relaxation time~\cite{shk08}.
Whereas Eqs.~\eqref{F23}-\eqref{F37} are written for any arbitrary angles $\alpha$, in order to get a more simple and clear
physical interpretation of the obtained results, below we put emphasis on the case when $\alpha=0^{\circ}$, i.e., when both the
currents flow along the WPP channels provoking the vortex movement perpendicular to them. As a result, below and throughout the
paper we consider only the $y$ components for both ratchet, dc and ac responses, omitting the index $y$ and $\langle\dots\rangle$
to simplify the notation.

The single vortex approximation used in Ref.~\cite{shk08} supposes the WPP period, $a$, to be large in comparison with the effective
magnetic field penetration depth, $\lambda$, and the temperature low enough to prevent smearing of singularities in the ratchet responses.
To accomplish this, until not stated otherwise, all the figure data are calculated for the dimensionless inverse temperature
$g\equiv U_p/2T=100$~\cite{shk08} representing a reasonable value, experimentally achievable, e.g., for thin Nb films either
grown on facetted sapphire substrates~\cite{sor07}, or furnished with nano-fabricated PPP landscapes~\cite{dob10-1,dob10-2},
where $U_p\simeq1000\div5000$~K and $T\approx8$~K. Assuming a triangular vortex lattice matching the PPP landscape at a
magnetic field $B\approx10$~mT, the pinning structure's
period is $a\approx400$~nm~\cite{phc11}. For a Nb film with a thickness $d\approx60$~nm, $\lambda(0)\simeq100$~nm~\cite{gub05}
(depends on temperature and the film's quality), so that the condition $d<\lambda<a$ can be experimentally satisfied.

Two groups of our findings commented below refer to the amplitude and frequency
dependencies of the ratchet responses $E^d$ and $\mathcal{\bar{P}}$. The dependencies are complementary to each other allowing
one to describe quantitatively $E^d$ and $\mathcal{\bar{P}}$ in the whole ($\xi^a, \Omega, \xi^d$)-space, as detailed next.

\subsection{Electric field $dc$ response}

One of the main questions in the study of the function $E^d(\xi^a|\xi^d,\Omega)$ given by Eq.~\eqref{F23} relies upon
the determination of the frequency and dc bias dependences of the ac amplitude threshold value, $\xi^a_c(\Omega,\xi^d)$,
which can be considered as an ac critical current magnitude for the dc ratchet response $E^d$ such that $E^d=0$ for $\xi^a<\xi^a_c$.
To accomplish this, we begin the graphical analysis with the ac amplitude dependence of the dc ratchet response considering specific
features in $E^d(\xi^a| \Omega, \xi^d)$ at low ($\Omega=0.01$), intermediate ($\Omega=1$), and high ($\Omega=3$) frequencies for
similar tilts, as depicted in Fig.~3.

Consider at first the curves in Fig.~3.a with $\Omega=0.01$ which corresponds to the adiabatic ratchet response. At large $\xi^a$,
for all the curves $E^d(\xi^a| \xi^d, \Omega\ll1)\simeq\xi^d$. These values will be explained in Sec.~IV in a rather simple
manner as this asymptotic behavior follows from Eq.~\eqref{adiabat9} for $E^+(\xi^d+\xi^a\cos\omega t)$ at $\xi^a\rightarrow\infty$
with ($+$) denoting the even component of $E$ regarding the change $\xi^a\rightarrow-\xi^a$. At small $\xi^a$ values, we observe
different behavior for curves with $\xi^d>1$ and $\xi^d<1$, namely for $\xi^d=1.05$ and $\xi^d=1.2$ the ratchet response is a threshold-free
one, whereas a threshold value, $\xi^a_c$, separates the non-dissipative and dissipative states at $\xi^d<1$. The magnitude of
the threshold is a decreasing function of $\xi^d$ and, in fact, is equal to $\xi^a_c=1-\xi^d$ which is evident for the adiabatic case.
The physical reason of the above difference follows from the fact that at $\xi^a=0$ and $\xi^d>1$ the vortex is in the
\emph{running} state with a slightly oscillating instantaneous velocity $d\mathrm{x}/dt$ and thus, nonzero electric field $E$,
whereas for $\xi^d<1$ the vortex is localized in one of the WPP wells.

Transferring from low ($\Omega=0.01$) to intermediate ($\Omega=1$) frequencies, several new distinctive features appear in Fig.~3.b
in comparison with the adiabatic case. First, the threshold values $\xi^a_c(\Omega=1,\xi^d)$ are larger than those ensuing
for $\Omega=0.01$ at similar subcritical dc tilts, i.e., $\xi^d<1$. To illustrate this in detail, we plot the $\xi^a_c(\Omega)$
dependence in Fig.~4 for a set of biases. All the curves demonstrate qualitatively similar behavior, i.e., a zero plateau
at $\xi^a<\xi^a_c$, a linear dependence at large $\xi^a>\xi^a_c$, and a nonlinear transition in between at $\xi^a\gtrsim\xi^a_c$.
These segments correspond to the adiabatic, intermediate, and high-frequency modes which are roughly separated by the straight lines
$\Omega\simeq0.1$ and $\Omega\simeq1$, respectively. It should be noted, that the curves $\xi^a_c(\Omega)$ in Fig.~4, calculated in
the present work for the dc-tilted cosine pinning potential, are qualitatively similar to those obtained experimentally on superconducting
Pb films with a non-tilted ratchet PPP (see Ref.~\cite{jin10} and Fig.~5 therein). The transition frequency from the adiabatic to nonadiabatic
case has been found at about 1~MHz for that system~\cite{jin10}.
\begin{figure}[t]
   \epsfig{file=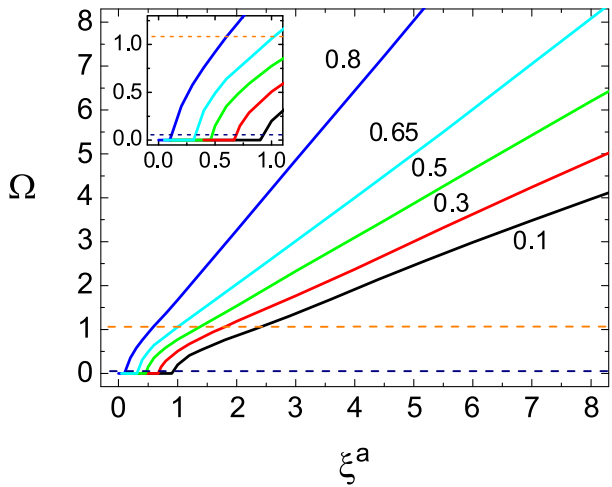,width=8cm}
   \caption{The frequency dependence of $\xi^a_c$ for the dc-tilted cosine pinning potential at different biases $\xi^d=0.1; 0.3; 0.5; 0.65; 0.8;$,
   as indicated. The navy and orange (online) dash lines represent rough separations between low ($\Omega\ll1$), intermediate ($\Omega\sim1$),
   and high frequency ($\Omega\gg1$) regimes. Inset: the nonlinear transition from the adiabatic to high-frequency regime in detail.
   The curves behave qualitatively similar to those obtained experimentally in Ref.~\cite{jin10}
   on superconducting Pb films with a non-tilted ratchet pinning potential.}
\end{figure}

Second, a difference in the $E^d(\xi^a|\xi^d,\Omega)$ behavior appears between $0.4\lesssim\xi^d_{middle}\lesssim0.7$, which looks
like damped oscillating curves, and the curves at $\xi^d\lesssim0.4$ and $\xi^d\gtrsim0.7$, which look like curves with phase-locked
regions (steps) in $\xi^a$. Whereas at small $\xi^d$ phase-locked regions ensue at $E^d=0$, at strong biases $\xi^d\gtrsim0.7$ these
flat segments appear at $E^d=1$. In Sec.~IV this will be discussed in detail within an approximate Bessel-function approach, originally
presented in Refs.~\cite{kau96,sha64} and used later~\cite{junxx} for $\Omega=1$. In fact, the authors of Ref.~\cite{junxx}
numerically calculated the Langevin equation in order to obtain the so-called dynamical current-voltage characteristics (CVCs)
of the resistively shunted Josephson junction model, which is equivalent to the model used here to analytically derive the dc
ratchet-response solution $E^d(\xi^a|\xi^d,\Omega)$ (notice the curve $E^d(\xi^a|\xi^d=0.8, \Omega=1)$
in Fig.~7.a of Ref.~\cite{junxx}). In our model, the corresponding Langevin equation with noise has been exactly solved in Ref.~\cite{shk08}
(see Eq.~(9) therein) and will be discussed in the noise-less limit (Eq.~\eqref{Q1} in Sec.~III) to clarify the origin of the steps in Fig.~3.b.

Lastly, consider the curves in Fig.~3.c at $\Omega=3$. From the above discussion it is clear that for $\xi^d>1$, the ratchet
responses $E^d$ are continuously oscillating curves without thresholds. On the contrary, for $\xi^d<1$ the responses have
thresholds whose magnitudes decrease with increasing $\xi^d$. As in the previous case, an interesting property of the dependence
$E^d(\xi^a|\xi^d,\Omega)$ in Figs.~2.b,c is the possibility for $E^d$ to decrease periodically (sometimes down to zero) with
increase of the driving amplitude $\xi^a$. Such a behavior of $E^d$ is in contrast to the behavior of the usual
dc-driven CVCs, even though in the presence of $\xi^a$, as these are always increasing functions of $\xi^d$.

Now we turn to the analysis of the frequency dependences of the ratchet response $E^d$ taken at fixed $\xi^a$ and $\xi^d$, as
represented in Fig.~5 for small ($\xi^a=0.5$), intermediate ($\xi^a=1$) and strong ($\xi^a=3$) ac amplitudes.
\begin{figure*}
   \epsfig{file=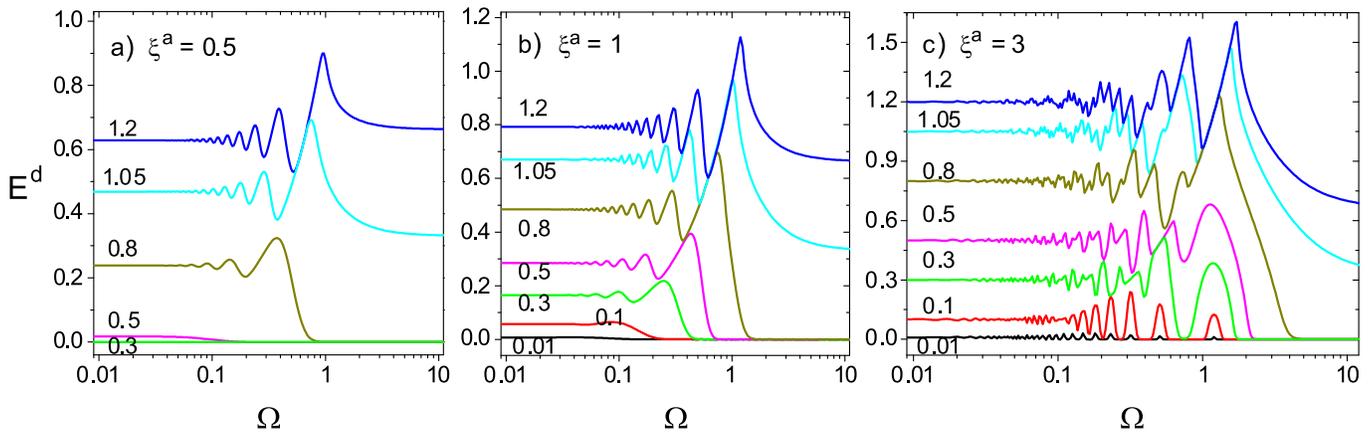,width=18cm}
   \caption{The voltage $E^d$  versus $\Omega$ for a set of biases $\xi^d=0.01; 0.1; 0.3; 0.5; 0.8; 1.05; 1.2$, as indicated,
   at small (a), intermediate (b) and strong (c) ac drives.}
\end{figure*}

At small ac drives $\xi^a=0.5$ (see Fig.~5.a) the curves vanish regardless of the frequency at small tilts $\xi^d<0.5$. This behavior
is evident, since if both the tilt value and ac drive amplitude are small, the vortices are localized at the bottoms of the WPP wells
which results in a non-dissipative state. With the gradual increase of the bias, for determinacy from $\xi^d=0.5$ to $\xi^d=0.8$,
the situation changes dramatically. At low frequencies the voltage drop gets substantially higher, whereas a zero-voltage tail spreads
over the high-frequency range. The former is a consequence of the \emph{running} vortex state, whereas the latter is a clear signature
of the \emph{localized} vortex state. These regions are separated by a threshold frequency $\Omega_c\equiv\Omega(\xi^d,\xi^a_c)$, which is in fact
already depicted in Fig.~4 and, as evident from the latter plot, is strongly dependent on both, $\xi^d$ and $\xi^a$ (compare with the
curves in Fig.~5.b for $\xi^a=1$). The tilt, $\xi^d$, determines the asymmetry of the WPP and the time needed for a vortex to get
from one to the next WPP well, whereas $\xi^a$ represents the ac driving force for a vortex which also competes with the height of
the initially symmetric WPP. This physically means that, if the ac driving frequency $\Omega$ is much less than the depinning frequency
$\omega_p\sim\hat{\tau}^{-1}$, the running state of
the vortex appears and it can visit several potential wells during the ac period. For a fixed ac amplitude $\xi^a$ and frequency $\Omega$,
the number of visited wells increases strongly with the increase of the tilt, thus resulting in a shift of the threshold frequency
towards higher $\Omega$. Another interesting feature in the $E^d(\Omega)$ curves appears as a maximum at $\Omega\simeq1$. Its magnitude
increases with increase of the frequency.

The  behavior of $E^d(\Omega| \xi^d, \xi^a)$ plotted for $\xi^a=3$ in Fig.~5.c, as representative for strong ac drives, can be
summarized as follows. In the adiabatic limit, when $\Omega\ll1$, the function $E^d(\Omega| \xi^d, \xi^a)$ coincides with the tilt
values $\xi^d$. At high frequencies $\Omega>1$, the curves either attenuate rapidly for subcritical tilts $\xi^d<1$ or approach
a constant value for $\xi^d>1$. In the intermediate regime, when $\Omega\simeq1$, the curves oscillate until the maximum is reached,
followed by the subsequent rapid decrease of the function $E^d(\Omega| \xi^d, \xi^a)$. Physically, this corresponds to the
pronounced reduction of the time $T_{\omega}/2$ with $T_{\omega}$ being the ac period, over which the moving force on the vortex keeps
its direction. As a result, the vortex may not longer visit several WPP wells, since the vortex displacement during $T_{\omega}$
is smaller than the WPP period $a$, even at strong ac amplitudes $\xi^a$.

Summarizing, the calculated ratchet behavior of $E^d(\xi^a|\xi^d,\Omega)$ differs substantially for a wide range of frequencies
$\Omega$, ac amplitudes $\xi^a$ and dc biases $\xi^d$. Throughout the frequency range this include a cut-off filter behavior of
the function $E^d(\xi^a|\xi^d,\Omega)$ with decreasing $\xi^a$ and increasing $\Omega$. In addition, at $\xi^d\lesssim0.4$
and $\xi^d\gtrsim0.7$ the curves demonstrate phase-locked peculiarities reminiscent of Shapiro steps~\cite{shk08} as well as a damped oscillatory
behavior at $0.4\lesssim\xi^d_{middle}\lesssim0.7$.

\subsection{Power absorption in $ac$ response}

In this subsection we first consider the behavior of the absorbed ac power $\mathcal{\bar{P}}$, calculated as function of the dimensionless
dc density $\xi^d$, ac density $\xi^a$, and frequency $\Omega$ in Sec.~V.E of Ref.~\cite{shk08}. Then we consider the graphs for
$E^d(\xi^a|\xi^d,\Omega)$ and $\mathrm{Re} Z_1(\xi^a|\xi^d,\Omega)$ in comparison with each other.
\begin{figure*}[t]
   \epsfig{file=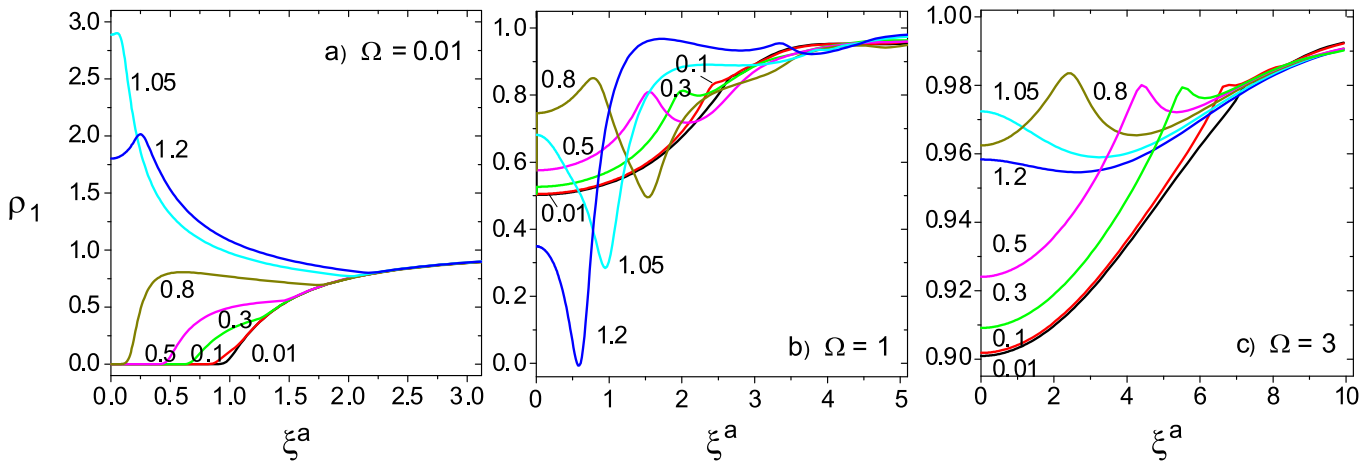,width=18cm}
   \caption{$\rho_1$ versus $\xi^a$ for a set of frequencies $\Omega$ and biases $\xi^d=0.01; 0.1; 0.3; 0.5; 0.8; 1.05; 1.2$, as indicated,
   in adiabatic (a), intermediate (b) and high-frequency (c) regime.}
\end{figure*}

Before entering the discussion it is useful to remember~\cite{shk08} that $\Omega=\omega\hat{\tau}$, where $\hat{\tau}$ is the relaxation
time and $\omega_p\sim\hat{\tau}^{-1}$ is the depinning frequency calculated in the linear ac approximation at $\xi^d=0$. We have to
point out that the bias dependences for the critical frequency $\Omega_c$ introduced in the previous subsection and the depinning
frequency $\omega_p$ are qualitatively opposite. Whereas the frequency dependence of the first Shapiro-like anomaly in $E^d(\xi^d)$
is given by the well-known expression $\Omega=\sqrt{(\xi^d)^2-1}$~\cite{lik86} for overcritical biases, the dependence for the
depinning frequency $\omega_p(\xi^d)=\omega_p\sqrt{1-(\xi^d)^2}$ has been obtained in the linear ac approximation at $\xi^a\rightarrow0$
and for subcritical tilts at $T=0$ only recently~\cite{shk10}. Thereby, in Ref.~\cite{shk10} it was shown  that $\omega_p(\xi^d)$ decreases with
increasing $\xi^d$ at $0<\xi^d<1$. The physical meaning of $\omega_p(\xi^d)$ at low temperature ($g\gg1$), i.e., when the creep factor $\nu_{00}$
(see Eq.~(107) in~\cite{shk08}) is small, follows from the fact that for low frequencies $\omega\ll\omega_p(\xi^d)$ (or $\Omega\ll1$)
pinning forces dominate and the vortex resistivity response $\rho_v$, being proportional to the absorbed power $\mathcal{\bar{P}}$, is
nondissipative (see Eq.~(2) in Ref.~\cite{shk08}), whereas at high frequencies $\omega\gg\omega_p(\xi^d)$ (or $\Omega\gg1$) frictional
forces dominate and $\rho_v$ is dissipative. The power absorbed per unit volume and averaged over the period of an ac cycle $\mathcal{\bar{P}}(\omega)$
was calculated in~\cite{shk08} (see Eq.~(84) therein), and can be written as
\begin{equation}
    \label{P01}
    \mathcal{\bar{P}}(\omega)=(\rho_f/2)[(\xi^a_x)^2 +
    (\xi^a_y)^2\textrm{Re}Z_1(\omega)],
\end{equation}
where $Z_1(\xi^a,\xi^d,\Omega,\alpha, g)$ is the nonlinear frequency- and dc and ac amplitude-dependent impedance. In Ref.~\cite{shk08}
it was shown that $Z_1$ plays the same role for the ac response as $\nu_0^{\omega}$ for the dc response.

Proceeding now to the analysis of the dependences $\mathcal{P}(\xi^a|\xi^d,\Omega)$, let us recall that to accomplish this
in the case $\alpha=0^{\circ}$, it is sufficient to calculate the ac resistivity $\rho_1\equiv \mathrm{Re}Z_1(\xi^a|\xi^d,\Omega)$
as function of its parameters, as presented in Figs.~5 and 6.

As for the dependences $E^d(\xi^a|\xi^d,\Omega)$ depicted in Fig.~4 and corresponding to low, intermediate, and high frequencies, the
curves $\rho_1(\xi^a|\xi^d,\Omega)$ are plotted at the same $\xi^d$ values. Consider at first the curves $\rho_1(\xi^a|\xi^d,\Omega)$
in Fig.~6.a at $\Omega=0.01$ which correspond to the adiabatic case $\omega\ll\omega_p(\xi^a)$. Here localized vortex states ensue at
$\xi^d<1$, and running states appear at $\xi^d>1$. For small tilts, $\xi^d<1$, an absorption threshold appears in the
$\rho_1(\xi^a|\xi^d,\Omega=0.01)$ curves at $\xi^a=\xi^a_c$. Here $\xi_c^a$ coincides with the critical ac magnitude for the corresponding
curves $E^d(\xi^a| \xi^d<1, \Omega=0.01)$. The physics of this threshold was earlier discussed for $\Omega=0.1$
(see Ref.~\cite{shk08}, Fig.~6) and for our adiabatic case ($\Omega=0.01$) may be connected with the $\omega_p(\xi^d)$ dependence.
At subcritical ac drives and overcritical biases, e.g., at $\xi^d=1.05$, as a consequence of the running vortex state, $\rho_1$
acquires large values. With the gradual increase of $\xi^a$ the curves exhibit a weak minimum and finally approach unity.

In the case of intermediate frequency (see Fig.~6.b for $\Omega=1$) the curves $\rho_1(\xi^a|\xi^d,\Omega)$ start from a nonzero
value regardless the bias $\xi^d$. A different behavior at subcritical and overcritical
tilts should be noted. At overcritical biases the running state for the vortex appears and the response is a consequence
of this motion with a slightly oscillating instantaneous velocity. On the contrary, at subcritical tilts the vortex is in the
oscillating state and if the frequency of the external excitation exceeds the depinning frequency $\omega_p(\xi^d)$, a nonzero
response appears as a result of the averaging of this oscillations. The bias dependence at $\xi^d<1$ in the limit of small
ac drives and $\Omega\ll1$ has been discussed above. Proceeding with the analysis of the curves shown in Fig.~6.b it
should be noted that at small biases $\xi^d\lesssim0.5$ the curves
increase monotonically, whereas at sufficiently large tilt values, e.g., at $\xi^d=0.8; 1.05; 1.2$ a pronounced minimum appears.
At overcritical biases this can lead to a sign change in $\rho_1(\xi^a|\xi^d,\Omega)$ in both $\xi^a$ and $\xi^d$. This evidence
is represented in detail in Fig.~7 overcritical tilts and weak ac drives. Figure 7 illustrates the presence of the
singular point $\xi^d=\sqrt{1+\Omega^2}$ in $\rho_1(\xi^d,\xi^a)$ at nonzero both the ac current amplitude and the temperature. A similar
singularity in the Josephson junction impedance problem has been discussed previously in the limit of small ac amplitude at zero
temperature~\cite{cof93}. Here we show that such a singularity does not vanish at nonzero temperature in the presence
of subcritical ac amplitude $\xi^a>0$. This evidence was left out of the scope of that work~\cite{cof93}.
Proceeding with the description of this anomaly it should be noted, that the dependence of $\rho_1(\xi^a|\xi^d,\Omega)$ on $\xi^a$ is less sharp
than in $\xi^d$. As far as high temperatures smear the singularity in both $\xi^a$ and $\xi^d$, the minimum can be more pronounced
if considered at lower temperatures ($g>100$).
\begin{figure}[b]
   \epsfig{file=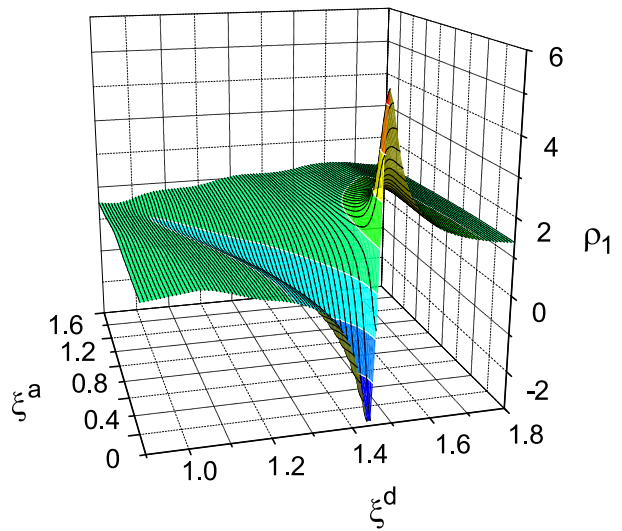,width=8cm}
   \caption{The ac resistivity $\rho_1$ versus $\xi^a$ and $\xi^d$ at $\Omega=1$ demonstrating a sharp singularity in $\xi^d$.
   This results in a sign change in the function $\rho_1(\xi^d,\Omega)$ at overcritical tilts $\xi^d\geq1$.
   The position of the minimum can be quantitatively calculated as $\xi^d=\sqrt{1+\Omega^2}$.}
\end{figure}
\begin{figure*}[t]
   \epsfig{file=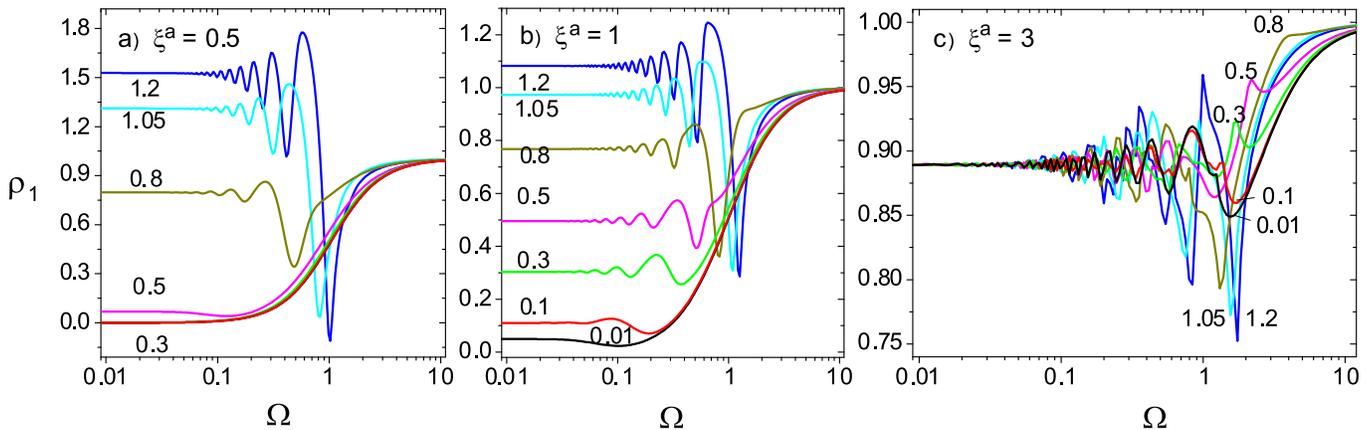,width=18cm}
   \caption{$\rho_1$ versus $\Omega$  for a set of biases $\xi^d=0.01; 0.1; 0.3; 0.5; 0.8; 1.05; 1.2$, as indicated,
   at small (a), intermediate (b) and strong (c) ac drives.}
\end{figure*}

Considering $\rho_1(\xi^a|\xi^d,\Omega)$ at high frequencies in Fig.~6.c, it is evident that $\rho_1(\xi^a|\xi^d,\Omega)$ approaches
unity even at small $\xi^a$ which again corresponds to the case $\omega\gtrsim\omega_p(\xi^d)$.

We go on with the analysis of the frequency dependence of $\rho_1$, represented in Fig.~8 for small $\xi^a=0.5$, intermediate $\xi^a\simeq1$
and strong $\xi^a=3$ ac drives for the same set of biases $\xi^d$.

At small ac drives (see Fig.~8.a) the curves demonstrate either monotonic behavior for $\xi^d\lesssim0.5$ or pronounced non-monotonic
behavior for $\xi^d\gtrsim0.5$. The monotonic curves at $\xi^d=0$ agree with the results of Coffey and Clem~\cite{cof91} who calculated
in linear approximation in $\xi^a$ the temperature dependence of the depinning frequency in a nontilted cosine pinning potential. In contrast to this monotonic behavior,
the nonmonotonic curves ($\xi^d\gtrsim0.5$) demonstrate two characteristic features. First, a pronounced power absorption in the adiabatic
regime. Second, a deep minimum in the power absorption at $\Omega\simeq1$. The appearance of this frequency- and temperature dependent
minimum was discussed in more detail in Ref.~\cite{shk08}. With the gradual increase of $\xi^a$, the value $\rho_1$ at low frequencies
remains the same, whereas the minimum shifts towards higher frequencies (see Fig.~8.b). In addition, at frequencies $\Omega\simeq0.5$
peculiarities in the curves become more pronounced. These can be smeared in turn when considered at higher temperatures ($g<<100$).

At strong ac drives, as represented in Fig.~8.c for $\xi^a=3$, already at very low frequencies $\Omega\ll1$ all the curves $\rho_1$
acquire large values and approach unity at high frequencies $\Omega\gg1$. Even though peculiarities in the dependence $\rho_1(\Omega)$
seem to be pronounced in the intermediate frequency range, they are sufficiently weak in comparison with those in Fig.~8.b when $\xi^a=1$.
A further increase of $\xi^a$ leads to slightly distorted curves throughout the frequency range.

The main results of this subsection can be summarized as follows. The power absorption in the ac response has been considered in
terms of the ac resistivity $\rho_1$ as a function of its driving parameters, $\xi^a$, $\Omega$, and $\xi^d$. While in the limiting cases
of small ac current in the absence of dc bias the well-known results of Coffey and Clem follow~\cite{cof91} and at strong dc biases a large power
absorption results, in agreement with the curves reported previously~\cite{shk08}, the appearance of a sign change in $\rho_1(\xi^d,\Omega)$
at a certain range of ac drives $\xi^a$ at overcritical biases $\xi^d$ is predicted for the first time in the present work.

\section{Discussion}
In this section, two physically different limiting cases at low and high frequencies will be considered at zero temperature
(i.e. $g\rightarrow\infty$) to augment the previous analysis with a more intuitive and visual interpretation.

The first case we consider is the adiabatic regime with $\Omega\ll1$. To discuss the dc ratchet response $E$ in this limit,
we employ static CVCs (see Eqs.~\eqref{adiabat3} and~\eqref{adiabat4} below) with $j=\xi^a\cos\mathbf{\omega} t, j_0=\xi^d$
and average $E^+(j+j_0)$ over the driving period $T_{\omega}=2\pi/\omega$. In this limit ($\omega\rightarrow0$),
while $\xi^a\gtrsim1$ the vortex may visit many potential wells of the WPP during the time $T_{\omega}/2$, i.e., when the
moving force on the vortex keeps its direction.
\begin{figure*}
   \epsfig{file=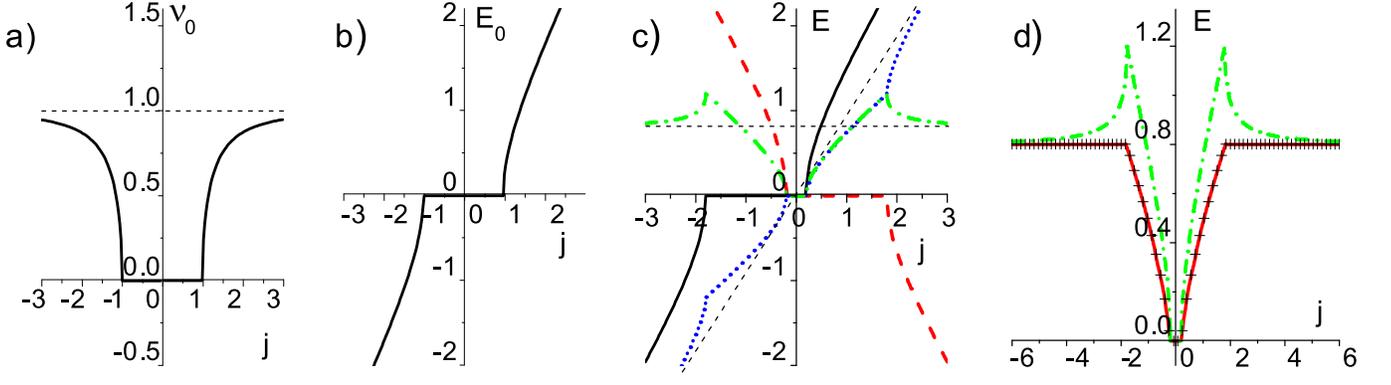,width=18cm}
   \caption{The functions $\nu_0(j)$ (a) and $E_0(j)$ (b) in the absence of a tilt in the adiabatic case at zero temperature.
   (c) In the presence of a dc bias $j_0=0.8$,  $E(j+j_0)$ (solid) and $E(-j+j_0)$ (dash) result in the appearance
   of even $E^+(j+j_0)$~(dash dot) and odd $E^-(j+j_0)$~(dot) components (see Eq.~\eqref{adiabat6}). The horizontal
   and diagonal thin dot lines represent asymptotics $E=0.8$ and $E=j$, respectively.
   (d) $E^+(j+j_0)$ (dash dot) is plotted together with the ratchet response $E^r$ (crosses, Eq.~\eqref{int2}) and
   the exact solution for $E^d(j, \Omega=0.001, j_0=0.8, g=1000)$ (solid, Eq.~\eqref{F23}).}
\end{figure*}

The second case is the non-adiabatic case with $\Omega\gg1$. In this limit, the vortex displacement during the time $T_{\omega}$
may be smaller than the WPP period $a$, even at strong ac densities $\xi^a$ and the Langevin equation for a vortex can
be solved in terms of the Bessel functions as detailed below.

\subsubsection{Adiabatic case}
We consider the vortex motion in a WPP, tilted by a dc bias $\xi^d<1$, and subjected to an ac drive with $\omega\rightarrow0$.
Our objective is to qualitatively point out why a rectified dc voltage appears in response to the ac input $\xi^a$.
To accomplish this, we first consider the dc CVCs for the cosine WPP (see Eqs.~\eqref{F23}) in the absence of an initial tilt, viz.,
\begin{equation}
    \label{adiabat1}
    E_0(j)=j\nu_0(j),
\end{equation}
where $-\infty<j<+\infty$ and
\begin{equation}
    \label{adiabat2}
    \nu_0(j)= \left\{
        \begin{array}{l}
        \sqrt{1-1/j^2},\qquad\qquad|j|>1,\\
        0,\qquad\qquad\qquad\qquad\quad |j|<1,\\
        \end{array}
        \right.
 \end{equation}
is the well-known nonlinear vortex mobility under the influence of the dimensionless generalized moving force $j$ in the $x$-direction
(see Eq.~52 in Ref.~\cite{shk08}). From Eq.~\eqref{adiabat2} follows, that $\nu_0(j)$ is an even function of $j$, i.~e. $\nu_0(j)=\nu_0(-j)$,
whereas $E_0(j)$ is an odd function of $j$, i.~e. $E_0(j)=-E_0(-j)$. The functions $\nu_0(j)$ and $E_0(j)$ are shown in Fig.~9.a and Fig.~9.b,
respectively.

As the cosine WPP is symmetric, i.~e. $U_p(x)=U_p(-x)$, it can establish ratchet properties only when tilted, i.e., we change
$j\rightarrow j+j_0$, where $j_0$ is the tilting dc bias. It is easy to see that depending on the sign of $j_0$, the tilt leads
to the shift of $\nu_0$ and $E_0$ along the $j$-axis by the value $|j_0|$ to the left (for $j_0>0$) or to the right (for $j_0<0$).
This is illustrated in Fig.~9.c where $j_0=0.8$ for definiteness. In this case
\begin{equation}
    \label{adiabat3}
    E_0(j)\rightarrow E(j+j_0)=(j+j_0)\nu(j+j_0),
\end{equation}
where
\begin{equation}
    \label{adiabat4}
    \nu_0(j+j_0)= \left\{
        \begin{array}{l}
        \sqrt{1-1/(j+j_0)^2},\qquad\qquad\quad j>1-j_0,\\
        0,\qquad\qquad\qquad\qquad -1-j_0<j<1-j_0,\\
        \sqrt{1-1/(j+j_0)^2},\qquad\quad j<(-1-j_0),\\
        \end{array}
        \right.
 \end{equation}

Whereas $E_0(j)$ and $\nu_0(j)$ are odd and even functions of $j$, respectively (see Fig.~9), from Eqs.~\eqref{adiabat3} and~\eqref{adiabat4}
follows that in the presence of a tilt $E(j+j_0)$ and $\nu(j+j_0)$ are neither even, nor odd in $j$ for $j_0\neq0$. In the following
it is suitable to present $E(j\pm j_0)$ as
\begin{equation}
    \label{adiabat5}
    E(j\pm j_0)=E^+(j\pm j_0)+E^-(j\pm j_0),
\end{equation}
where
\begin{equation}
    \label{adiabat6}
    E^{\pm}(j\pm j_0)=[E(j\pm j_0)\pm E(-j\pm j_0)]/2
\end{equation}
and
\begin{equation}
    \label{adiabat7}
    E^{\pm}(j\pm j_0)=\pm E^{\pm}(-j\mp j_0)
\end{equation}
are the even and odd parts of $E(j\pm j_0)$ with respect to change $j\rightarrow-j$.

On the other hand, it is clear that
\begin{equation}
    \label{adiabat8}
    E(j\pm j_0)=-E(j\mp j_0),
\end{equation}
because when we change the sign of the full current [i.~e. ($j\pm j_0)\rightarrow(-j\mp j_0)$], then $E$ will be an odd function of
the full current. If we then apply the $\pm$-operation to $E$ given by Eq.~\eqref{adiabat8} and take into account Eq.~\eqref{adiabat6},
we arrive at the important conclusion that
\begin{equation}
    \label{adiabat9}
    E^{\pm}(j\pm j_0)=\mp E^{\pm}(j\mp j_0).
\end{equation}
From Eq.~\eqref{adiabat9} it follows that $E^+(j+j_0)=-E^+(j-j_0)$.  This means $E^+(j+j_0)$ changes its sign when changing the
sign of $j_0$, whereas $E^-(j-j_0)=E^-(j+j_0)$ does not change its sign. From the physical viewpoint it means
that $E^+(j+j_0)$, even in $j$ and odd in $j_0$, is responsible for the ratchet response, whereas $E^-(j+j_0)$, which is odd in $j$ and
even in $j_0$, describes the usual CVCs response, analogous at $j\gg1$ to that at $j_0=0$. Actually, if we take into account that
from Eq.~\eqref{adiabat4} follows $\lim_{j\rightarrow\infty}\nu(\pm j+j_0)=1$, then from Eq.~\eqref{adiabat7} at once follows that
\begin{equation}
    \label{adiabat10}
    \lim_{j\rightarrow\pm\infty}E^+(j+ j_0)=j_0,
\end{equation}
\begin{equation}
    \label{adiabat11}
     \lim_{j\rightarrow\pm\infty}E^-(j+ j_0)=j.
\end{equation}

At last, let us perform the change $j\rightarrow j\cos\omega t$ and consider the function $E(j\cos\omega t + j_0)$, since this represents
a more close correlation with the exact results obtained by using Eq.~\eqref{F23}. To derive the average dc ratchet solution $E^r$ in response
to to the input current $j\cos\omega t + j_0$, one needs to integrate the function $E$ over the ac current period $T_{\omega}$
\begin{equation}
    \label{int1}
    E^r\equiv\frac{1}{T_{\omega}}\int_{\substack{0}}^{\substack{T_{\omega}}}dt~E(j\cos\omega t + j_0).
\end{equation}
Equation~\eqref{int1} can be reduced to the sum of two integrals
\begin{equation}
    \label{int2}
    E^r\equiv\frac{1}{\pi}[\int_{\substack{0}}^{\substack{\pi/2}}d\varphi~E(j\cos\varphi + j_0) +
    \int_{\substack{0}}^{\substack{\pi/2}}d\varphi~E(-j\cos\varphi + j_0)],
\end{equation}
with the integrals to be taken over $\varphi=\omega t$ only where $\nu(j\cos\omega t + j_0)$
is nonzero in accordance with Eq.~\eqref{adiabat4}, i.e.,
\begin{equation}
    \label{int3}
    \varphi > \left\{
        \begin{array}{l}
         \arccos(\frac{1-j_0}{j}),\qquad\qquad j>1-j_0,\\
        \arccos(\frac{-1-j_0}{j}),\qquad\qquad j<-1-j_0,\\
        \end{array}
        \right.
 \end{equation}

Equation~\eqref{int2} can be rewritten in another equivalent form
\begin{equation}
    \label{int4}
    E^r\equiv\frac{2}{\pi}\int_{\substack{0}}^{\substack{\pi/2}}d\varphi~E^{+}(j\cos\varphi + j_0),
\end{equation}
which represents the even component of the dc ratchet response and will be compared next with the temperature-dependent response $E^d(\xi^a)$.

We now take a closer look at Figure~9 considering firstly the curves $E(j+j_0)$ (solid) and $E(-j+j_0)$ (dash) in Fig.~9.c.
As it follows from Eq.~\eqref{adiabat10}, $E(j+j_0)$ is zero at $-1-j_0<j<1-j_0$
as far as $E(-j+j_0)$ vanishes at $-1+j_0<j<1+j_0$. A rapid increase in both the functions in $j$ in the vicinity
to $\pm(1-j_0)$ and $\pm(1+j_0)$ should be noted. In the adiabatic limit (see Eq.~\eqref{adiabat6}),
$E^+(j+j_0)$ and  $E^-(j+j_0)$ are shown in Fig.~9.c by dash dot and dot lines, respectively. They inherit both the
bump-like peculiarities in $E(\pm j+j_0)$ at $\pm(1+j_0)$.
Next we turn to Fig.~9.d where adiabatic (Eq.~\eqref{adiabat10}, dash dot), approximate (Eq.~\eqref{int2}, crosses),
and exact (Eq.~\eqref{F23}, solid) ratchet responses are shown together. The asymptotic behavior of all the curves is
in agreement with Eq.~\eqref{adiabat10}. Peculiarities are highly pronounced in the adiabatic
solution~\eqref{adiabat6}, calculated in fact in response to the square wave ac current.
In contrast to this, the approximate ratchet solution~\eqref{int2} for the cosine ac current practically coincides with the exact
solution~\eqref{F23} calculated in the limit of very small frequencies $\omega=0.001$ and very low temperature $g=1000$.

By this way, we conclude that the simple approach in terms of the static CVCs can explain qualitatively the form of the curves
$E^d(\xi^a)$ in Fig.~3.a in the adiabatic limit. A good quantitative agreement between the approximate and exact solutions
is revealed at very low frequencies and temperatures.

\subsubsection{Nonadiabatic case}
\begin{figure}[t]
   \epsfig{file=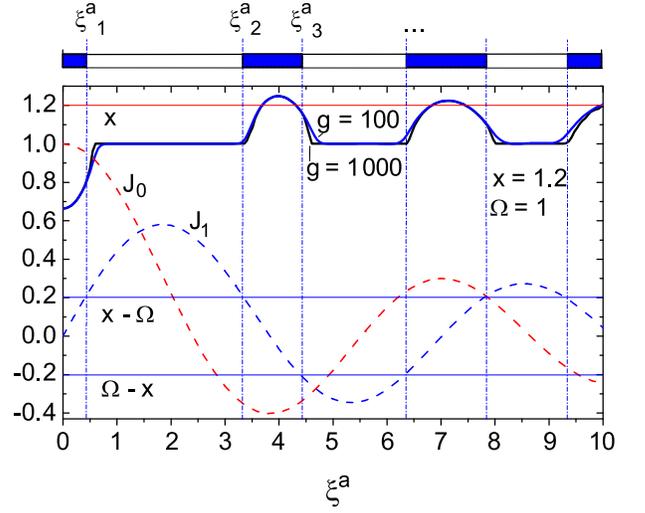,width=8cm}
   \caption{The functions $J_0(\xi^a)$ and $J_1(\xi^a)$ (dash) are plotted together with straight lines $\xi^d=1.2$ and
   $\xi^d-\Omega=\pm0.2$. The phase locked regions in $E^d(\xi^a | \Omega=1, \xi^d=1.2, g=100)$ (blue line) are connected by
   vertical lines for clarity. The light segments of the stripe on top of the plot indicate phase locked regions with $\xi^a_i$
   denoting the roots of Eq.~\eqref{Q6} for n=0 and n=1. A better fit of the noiseless approximate solution (23) can be achieved
   if the ratchet response~\eqref{F23} is calculated at very low temperatures, such as for $g=1000$ (black line).}
\end{figure}
In order to explain qualitatively the results of the exact calculations of $E^d(\xi^a|\xi^d,\Omega)$ at low temperatures
($g=100$), presented in Fig.~3b,c for intermediate and high frequencies ($\Omega\sim1$ and $\Omega\gg1$) of the $ac$ driving
force, we use a more simple approach which ignores the noise ($g\rightarrow\infty$). In this limit, the equation of motion for the
dimensionless vortex coordinate $\mathrm{x}$ (see Eq.~(9) in~\cite{shk08}) reduces to
\begin{equation}
    \label{Q1}
    d\mathrm{x}/dt + \sin \mathrm{x} = \xi^d + \xi^a\cos\Omega t
\end{equation}
which is analogous to the well-known equation of motion for the phase difference in the $ac$-driven resistively shunted
Josephson-junction model~\cite{lik86} at zero temperature. For high and intermediate frequencies ($\Omega\gg1$ and $\Omega\sim1$) and
for $\xi^d>0.5$ Eq.~\eqref{Q1} can be approximately analyzed and solved in the spirit of the ansatz of Ref.~\cite{kau96}.
In these limiting cases we simply assume that the velocity of the vortex is sinusoidal in accordance with
\begin{equation}
    \label{Q2}
    d\mathrm{x}/dt = \langle d\mathrm{x}/dt\rangle + \xi^a\cos\Omega t
\end{equation}
and determine the constant $\langle d\mathrm{x}/dt\rangle$ by requiring that $d\mathrm{x}(t)/dt$ satisfies Eq.~\eqref{Q1}.
Having integrated Eq.~\eqref{Q2} we obtain
\begin{equation}
    \label{Q3}
    \mathrm{x}(t) = \mathrm{x}_0 + \langle d\mathrm{x}/dt\rangle t + (\xi^a/\Omega)\sin\Omega
    t,
\end{equation}
where $\mathrm{x}_0$ is a second constant to be determined. Substituting Eq.~\eqref{Q3} into Eq.~\eqref{Q1} and using the expansion of
$\sin\mathrm{x}$ in a harmonic series according to formulas originally suggested by Shapiro et. al. in Ref.~\cite{sha64}, i.e.,
\begin{equation}
    \label{Q4}
        \begin{array}{l}
        \cos(\mathrm{x}\sin\varphi)=\sum_{\substack{k=-\infty}}^{\substack{\infty}}J_k(\mathrm{x})\cos(k\varphi),\\
        \\
        \sin(\mathrm{x}\sin\varphi)=\sum_{\substack{k=-\infty}}^{\substack{\infty}}J_k(\mathrm{x})\sin(k\varphi)\\
        \end{array}
\end{equation}
where $J_k(\mathrm{x})$ is the $k$-th order Bessel function, we obtain
\begin{equation}
    \label{Q5}
    \langle d\mathrm{x}/dt\rangle  = \xi^d -\sum_{\substack{k=-\infty}}^{\substack{\infty}}J_k(\xi^d/\Omega)\sin[\mathrm{x}_0 + (\langle d\mathrm{x}/dt\rangle+k\Omega)t]
\end{equation}
from which $\langle d\mathrm{x}/dt\rangle$ can be found self-consistently. In Eq.~\eqref{Q5} $\mathrm{x}_0$ is an
arbitrary coordinate. For values $\langle d\mathrm{x}/dt\rangle$ that are not integral multiples of $\Omega$, the
term $\sin \mathrm{x}$ does not contribute a constant component to Eq.~\eqref{Q1}, and equating the constant terms yields (after averaging over
one period for locking into the $n$-th region)
\begin{equation}
    \label{Q6}
    \xi^d - n\Omega = J_n(\xi^a/\Omega)(-1)^n\sin\mathrm{x}_0.
\end{equation}
\begin{figure}[t]
   \epsfig{file=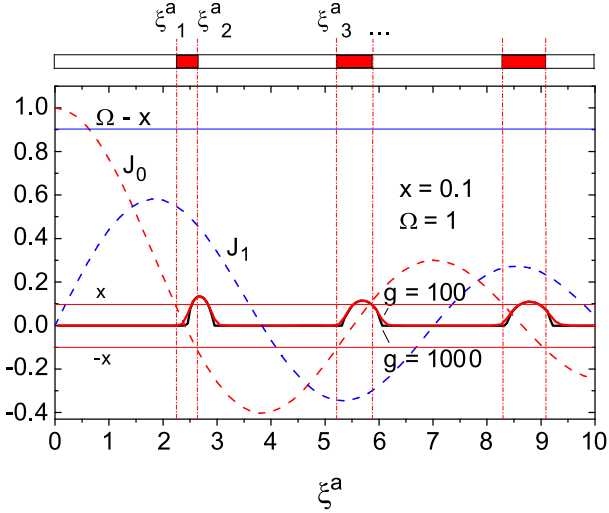,width=8cm}
   \caption{The functions $J_0(\xi^a)$ and $J_1(\xi^a)$ (dash) are plotted together with straight lines
   $\pm\xi^d(=0.1)$ and $\Omega-\xi^d=0.9$ to illustrate the phase locked regions in $E^d(\xi^a | \Omega=1)$ at $\xi^d=0.1$.
   Other details are similar to those in Fig.~10.}
\end{figure}

In order to use these results for the explanation of the behavior of $E^d(\xi^a|\xi^d, \Omega=1)$ (see Fig.~3.b) we consider two cases.
The first case considers $\xi^d=1.2$ for which at $\xi^a=0$ the vortex is in the running state. From Fig.~(9) we see, however, that for
$\xi^a_1<\xi^a<\xi^a_2$ the value of $E^d$ is locked into the phase of the ac periodic driving at $n=1$ (see Eq.~\eqref{Q6}). From Eq.~\eqref{Q6}
follows that for $n=0$ at $\xi^d=1.2$ and $\Omega=1$ this equation has no solution at any value of $\mathrm{x}_0$. For $n=1$ we have
\begin{equation}
    \label{Q7}
    J_1(\xi^a)\sin\mathrm{x}_0=-0.2
\end{equation}
at $\xi^a_i$ values which satisfy the condition $J_1(\xi^a)=\pm0.2$ and strictly correspond to the phase-locked regions for this curve.
Between these regions $E$ has a bump-like form with an increasing width and decreasing height with the increase of $\xi^a$.
For $n=2$ Eq.~\eqref{Q6} has no solution.

The second case deals with $\xi^d=0.1$, for which at $\xi^a<\xi^a_c$ the vortex is in the locked state with $n=0$. In this case the
equation $J_0(\xi^a)=\pm0.1$ has many solutions (see Fig.~11). Considering $n=1$ leads to the equation $J_1(\xi^a)\sin\mathrm{x}_0=\pm0.9$
which has no solution, however.

\section{Conclusion}
In this work we proposed an exactly solvable two-dimensional model structure for the study of the frequency-dependent ratchet effect
in superconducting film with a symmetric planar pinning potential, tilted by a dc bias, also known as a tilted ratchet.
We have theoretically examined the strongly nonlinear nonadiabatic tilted ratchet behavior of the two-dimensional vortex system
of a superconductor as a function of the (ac+dc) transport current density $\mathbf j$, the frequency $\omega$, and the
temperature $T$. The nonlinear (in $\mathbf j$) resistive behavior of the anisotropic vortex ensemble is be caused by the presence
of anisotropic pinning with the
symmetry of the PPP. It is physically obvious that such a pinning at low enough temperatures leads to anisotropy of the vortex
dynamics since it is much easier for vortices to move along the pinning channels (the guiding effect in the flux-flow regime,
which is linear in the current) than in the perpendicular direction, where it is necessary for them to overcome the pinning potential
barriers. The latter is also a source of nonlinearity of the dc+ac responses. If under variation of one of the "external"
driving parameters $\mathbf j$, $T$, and $\alpha$, the intensity of the manifestation of the indicated nonlinearity is weakened,
this weakening will lead to an "effective isotropization" of the vortex dynamics, i.e., to a convergence (and in the limit of the
absence of nonlinearity, to coincidence) of the directions of the mean velocity vector of the vortices and the Lorentz force~\cite{ShSo}.

It is physically clear that current, temperature, and angle $\alpha$ have qualitatively different effects on the weakening of
the pinning and the corresponding transition from anisotropic vortex dynamics to isotropic. With the growth of $\mathbf j$ the Lorentz
force $\mathbf{F}_{\emph{L}}$ grows and the height of the potential barrier decreases, so for $j \geqslant j_{cr1},j_{cr2}$ these
barriers essentially disappear. Here $j_{cr1,2}$ are the crossover currents for these transitions to occur regarding the right- and
left-hand PPP barriers. The quantities $j_{cr1,2}$ depend on $\alpha$ by virtue of the fact that the probability of overcoming
the barrier is governed not by the magnitude of the force $F_\emph{L}$, but only by its transverse component $F_\emph{L}\cos\alpha$,
so that $j_{cr1,2}(\alpha) =j_{cr1,2}(0)/\cos\alpha$ grows with increasing of $\alpha$. Since an increase in temperature $T$ always
increases the probability of overcoming the pinning barrier, the transition to isotropization of the vortex dynamics is much
steeper in $T$, the smaller the pinning barrier is. However, although general formulas for the ratchet responses
(see Eqs.~\eqref{F23}-\eqref{F37}) include both, the angle and temperature dependences, in the present work we
used $\alpha=0^{\circ}$ and $g\equiv U_p/2T=100$ for simplicity.

Proceeding now to a short description of the main theoretical results, we note here that an exact analytical representation of the
nonlinear ac-driven rachet response of the investigated system in terms of a matrix continued fraction was possible thanks to the use
of a simple but physically realistic model of anisotropic pinning with a tilted cosine WPP. The exact solution obtained made it
possible for the first time to consistently analyze not only the qualitatively clear vortex dynamics of the adiabatic ratchet
effect, but also the nontrivial ratchet behavior at intermediate and high frequencies of the ac drive. Below we turn to a short
presentation of our results taken in comparison with experimental results, presented recently in Ref.~\cite{jin10}.

First of all, simple inspection of our exact expressions~\eqref{F23} for the dc ratchet response shows that a magnetic field inversion
does not change the sign of $E^d(\xi^d|\xi^a, \omega)$, as on the right side of Eqs.~\eqref{F23} the $E^d_{x,y}$-components do not depend on the
index $n$, which determines the $\mathbf{B}$-inversion. On the other hand, we should point out that for the adiabatic ratchet studied in detail
in Ref.~\cite{shk09} for the asymmetric PPP without dc bias, the dc response changes its sign after $\mathbf{B}$-inversion
(see Eq.~(16) therein). Since a clear sign change in $V_{dc}$ has been experimentally observed at field inversion~\cite{jin10}, we
conclude that the dc ratchet response in that work should be described by a model with an asymmetry of the PPP~\cite{shk09}.

An another interesting difference between the tilted ratchet described in the present paper and the asymmetric ratchet without a tilt~\cite{shk09}
consists in their asymptotic behavior at $j\rightarrow\pm\infty$ in the adiabatic regime. In this limit, the tilted ratchet adiabatic response
is finite and equal to the tilt value, as one can see from Eq.~(16) at $j_0<1$, whereas for the asymmetric ratchet this response is zero
because $\lim_{j\rightarrow\pm\infty}\nu^-(j)\sim1/j^3\rightarrow0$~\cite{shk09}.

Comparing our frequency-dependent results with analogous experimental findings presented in Ref.~\cite{jin10} we should underline that
in spite of the simple WPP and the single-vortex approximation used in our theoretical model we, however, can qualitatively explain from
one and the same point of view main experimental results of that work. In particular, our expressions~\eqref{F23} and~\eqref{F64}
describe (i)~the critical ac current dependence in a wide frequency range covering the transition from adiabatic to nonadiabatic,
with both, a frequency-independent plateau at low frequencies, a direct dependence at high frequencies and a nonlinear transition in between,
(ii)~the appearance of phase-locking regions in the dependence of $V_{dc}$ on $I_{rf}$, (iii)~a weakening of the ratchet effect at extremely
high frequencies, and (iv) the possibility of a sign change for the absorbed power in ac response within a certain range of the driving parameters.
This is in contradistinction to different explanations used in~\cite{jin10}, which is a consequence of the absence of a well-defined
theoretical model due to the complexity of the PPP employed. In particular, due to this reason the authors~\cite{jin10} compelled to
employ different approaches for the explanation of their experimental results, such as introducing sometimes the vortex mass, or
appealing to the vortex-vortex interaction, or not taking into account the tilting parameter, or leaving uncommented the
high frequency power absorption behavior.

It should be stressed that the single vortex approximation used in this work may only be valid at small magnetic fields preferably less
than the first matching field, so no collective effects are captured in the model considered here. Whereas vortex ratchet reversals as a
function of field have been studied by authors in a number of simulations and experimental works and have been explaining as a result
of collective effects, such as vortex-vortex interactions~(see, e.g., Ref.~\cite{din09} and references therein), a remark on the extension
of our theoretical study is now to be made. As far as the model described in the present paper refers to the tilted-potential ratchet, a thorough
theoretical description of the rocking-ratchet response in superconducting films with an asymmetric WPP is currently under way and will
be reported in a forthcoming publication~\cite{shk12}. There will be shown that the single-vortex approximation can also
lead to the ratchet reversals, when we consider an asymmetric WPP in the presence of a tilting bias. Finally, we would like to stress,
that our exactly solvable single-vortex model explicitly shows that many important and interesting nonlinear ratchet effects, which can
be observed in particular at high frequencies, follow even from such a simple model for one vortex in periodic ratchet WPP.
Though experimental verification of the predictions of both the models can be performed, for instance, on Nb thin
films with nano-fabricated WPP landscapes~\cite{dob10-1,dob10-2}, the first portion of ratchet data still remains to be seen.

\section{Acknowledgements}
V.A. Shklovskij thanks the Deutsche Forschungsgemeinschaft (DFG) for financial support through Grant No. HU 752/7-1.
O.V. Dobrovolskiy gratefully acknowledges financial support by the DFG through Grant No. DO 1511/2-1.
The authors would like to thank M. Huth for the critical reading.

\end{document}